\documentclass[fleqn,12pt,twoside]{article}
\usepackage{espcrc1}

\usepackage{graphicx}
\usepackage[figuresright]{rotating}

\newcommand{\AmS}{{\protect\the\textfont2
  A\kern-.1667em\lower.5ex\hbox{M}\kern-.125emS}}

\title{Dirac's inspired point form and hadron form factors}

\author{B. Desplanques\address{Laboratoire de Physique Subatomique 
et de Cosmologie \\(UMR CNRS/IN2P3-UJF-INPG),  
F-38026 Grenoble Cedex, France}\thanks{desplanq@lpsc.in2p3.fr}   }

\begin{document}

\maketitle

\begin{abstract}
Noticing that the ``point-form'' approach referred to 
in many recent works implies physics described on hyperplanes, 
an  approach inspired from Dirac's one, which involves a hyperboloid surface, 
is presented. A few features pertinent to this new approach are emphasized. 
Consequences as for the calculation of form factors are discussed.
\end{abstract}

\section{INTRODUCTION}
Looking at  a  Hamiltonian formulation of relativistic dynamics, Dirac 
was led to consider various forms, depending on the symmetry properties 
of the hypersurface that is chosen in this order \cite{Dirac:1949cp}. 
Accordingly, the generators of the Poincar\'e algebra drop into dynamical 
or kinematic ones. Among the different forms, the point-form approach, 
which is based on a hyperboloid surface, $t^2-\vec{x}\,^2=\tau$, 
is probably the most aesthetic one in the sense that the space-time 
displacement operators, $P^{\mu}$,  are the only ones to contain 
the interaction while the boost and rotation operators, $M^{\mu\,\nu}$ 
altogether, have a kinematic character. This approach is also 
the less known one, perhaps because dealing with a hyperboloid surface 
is not so easy as working with the hyperplanes that underly the other 
forms (instant and front). It nevertheless received some attention 
recently within the framework of relativistic quantum mechanics (RQM). 
Due to the kinematic character of boosts, its application to the calculation 
of form factors can be easily performed {\it a priori} and, moreover, 
these quantities generally evidence the important property of being Lorentz 
invariant. 

A ``point-form'' (``P.F.'') approach has been successfully used 
for the calculation of the nucleon form factors \cite{Wagenbrunn:2000es}. 
It however fails in reproducing the form factor of much simpler systems, 
including the pion \cite{Desplanques:2001zw,Amghar:2002jx,Amghar:2003tx}. 
The asymptotic behavior is missed and the drop-off 
at small $Q^2$ is too fast in the case of a strongly-bound system. 
Analyzing the results, it was found that this ``point-form'', 
where the dynamical or kinematic character of the  Poincar\'e 
generators is the same as for Dirac's one, implies physics
described on hyperplanes \cite{Desplanques:2001ze}. This approach is 
nothing but that one presented by Bakamjian as being an ``instant form~$\cdots$ 
which displays the symmetry properties inherently present 
in the point form'' \cite{Bakamjian:1961}. Sokolov  mentioned it
was involving ``hyperplanes orthogonal $\cdots$ to the 4-velocity 
$\cdots$  of the system'' under consideration, adding it was not 
identical to the point form proposed by Dirac \cite{Sokolov:1985jv}. 
Developing an approach more in the spirit of the original one 
therefore remains to be made. 

In this contribution, we present an exploratory work that is motivated 
by Dirac's point form and, consequently, implies physics described on 
hyperboloid-type surfaces. Due to the lack of space, we only consider here 
the main points while details can be found in Ref.~\cite{Desplanques:2004rd}.
How this new approach does for hadron form factors 
is briefly mentioned. 
\section{FORMALISM: A FEW INGREDIENTS}
Each RQM approach is characterized by the relation that the 
momenta of a system and its constituents fulfill off-energy shell. This one 
is determined by the symmetry properties of the hypersurface which 
the physics is formulated on. In absence of particular direction 
on a hyperboloid-type surface, it necessarily takes the form of a
Lorentz scalar. One should also recover the momentum 
conservation in the non-relativistic limit, $\sum_i\vec{p_i}=\vec{P}$. 
Thus, for the two-body system we are considering here, the expected relation 
could read: 
\begin{equation} 
(p_1+p_2-P)^2=(\vec{p_1}+\vec{p_2}-\vec{P})^2-(e_1+e_2-\;E_P)^2=0\,.
\label{scalar}
\end{equation}
Such a constraint is obtained from integrating plane waves on the hypersurface, 
$x^2=0$:
\begin{equation} 
\int d^4x \;  e^{i(p-p') \cdot x}\; \delta(x^2)\;\epsilon(U \!\cdot\! x) 
=4i\pi^2\;\delta((p-p')^2)\; \epsilon(U\!\cdot\!(p-p'))\,,
\end{equation}
where $p^{\mu}$ and $p'^{\mu}$ are replaced by $(p_1\!+\!p_2)^{\mu}$ and 
$P^{\mu}$, and $U^{\mu}$ satisfies $U^2 \geq 0$. To understand the ingredients 
entering the l.h.s. of the above equation, the ``time'' evolution should be 
examined \cite{Desplanques:2004rd}. This goes beyond considering 
the upper part of a hyperboloid surface often mentioned in the literature. 
Interestingly, Eq. (\ref{scalar}) can be cast into the following form: 
\begin{equation} 
(\vec{p_1}+\vec{p_2}-\vec{P})= \hat{u}\;(e_1+e_2-\;E_P)\;
(\hat{u}^2=1, \; \hat{u} \; {\rm not \; fixed} )\,,
\end{equation}
which is very similar to a front-form one, but the unit vector, $\hat{u}$, 
has no fixed direction.

The next step consists in considering a wave equation, which can be obtained 
from taking the square of the momentum operator, $P^{\mu}$:
\begin{eqnarray}
\nonumber 
\Big(M^2-p^2\Big)\;\Phi_P(\vec{p}_1,\vec{p}_2)= - 
\int  \int 
\frac{d\vec{p}\,'_1}{(2\,\pi)^3} \; \frac{d\vec{p}\,'_2}{(2\,\pi)^3} \;\;    
\frac{1}{(2\,e_1 \;2\,e_2\;2\,e'_1\;2\,e'_2)^{1/2} }\; \hspace*{2cm} 
\nonumber \\   \times \, 
(p+p')\!\cdot\! \; \partial_{p-p'} \Bigg(
4\,\pi^2\; \delta \Big((p-p')^2\Big)\; \epsilon \Big(U \!\cdot\! (p-p') \Big) 
\Bigg)\; 
\frac{4\,m^2\;g^2}{\mu^2+\cdots} \;\; \Phi_P(\vec{p}\,'_1,\vec{p}\,'_2) \, ,
\end{eqnarray} 
where  $ \;\; p^{\mu}=p_1^{\mu}+p_2^{\mu}$. One should determine under which
conditions it admits solutions verifying Eq. (\ref{scalar}) and, at the same
time, leads to a relevant mass operator. With this aim, we assume 
$U^{\mu}\propto c(p-P)^{\mu}+c'(p'-P)^{\mu} $, from which we get:
\begin{equation}
\frac{\vec{U}}{U^0}=\hat{u}= \frac{\vec{p}-\vec{P}}{e-E_P}= 
\frac{\vec{p}\,'-\vec{P}}{e'-E_P}= 
\frac{\vec{p}\,''-\vec{P}}{e''-E_P}= \cdots \, .
\end{equation} 
This relation shows that the orientation of $\hat{u}$ is conserved, 
which greatly facilitates the search for a solution. While doing so, 
a Lorentz-type transformation adapted from the Bakamjian-Thomas one 
\cite{Bakamjian:1953kh} has to be made. The constituent momenta are expressed in terms 
of the total momentum, $\vec{P}$, and the internal variable, $\vec{k}$, 
while verifying Eq. (\ref{scalar}). Moreover, the interaction 
is assumed to fulfill constraints but these ones, which amount 
to take into account higher-order meson-exchange contributions, 
are actually well known as part of the general construction of the 
Poincar\'e algebra in RQM approaches \cite{Keister:sb}. 

The present point form implies that the system described in this way evidences a
new degree of freedom. In the c.m., a zero total momentum is obtained by adding
the individual contributions of constituents and an interaction one,
consistently with the fact that $\vec{P}$ contains the interaction. The
configuration so obtained points isotropically to all directions as sketched
in Fig. 2 of Ref.  \cite{Desplanques:2004rd}. This new degree of freedom appears
explicitly in the definition of the norm, beside the integration on the internal 
$\vec{k}$ variable:
\begin{equation}
N =\int \frac{d\vec{k}}{(2\pi)^3}\;\phi^2_0(k)  
\int \frac{d\vec{u}}{2\,\pi }\; \delta(1-\vec{u}\,^2)\;
\frac{M^2}{(u \!\cdot\! P)^2}\, ,
\end{equation}
where $\phi_0(k)$ represents a solution of a mass operator. 
Another aspect of the present point form concerns the velocity operator, 
$\vec{V}$, entering the construction of the Poincar\'e algebra, and the
corresponding $\vec{P}$.
Their expressions, which differ from  earlier ones, read:
\begin{equation}
\vec{V}= \frac{1}{M} \,
\Big(\vec{p}+ \vec{u}\; \frac{4\,m\;\tilde{V}}{2\,u \! \cdot 
\!P}\Big), \;
\vec{P}=\vec{p}+ \vec{u}\; \frac{4\,m\;\tilde{V}}{2\,u \! \cdot \!P}\;\;\;\;
\Big({\rm earlier\; ``P.F.":} \;\vec{V}= \frac{\vec{p}}{2\,e_k},\; 
\vec{P}=\frac{M}{2\,e_k}\;\vec{p}\Big)\, .
\label{velocity}
\end{equation}
Despite unusual features, the present point form can be
consistently developed. It evidences similarities with the instant 
and front forms in that the hypersurface it is formulated on is 
independent of the system under consideration, which is at the origin 
of the above constraints. These ones are absent in the earlier 
``point form'' where the kinematic character of boosts is trivial, 
the operation affecting both the system and the hyperplane used for its 
description, their respective velocity and orientation being related.
\section{SOLVED AND UNSOLVED PROBLEMS, DISCUSSION, OUTLOOK}
\begin{figure}[htb]
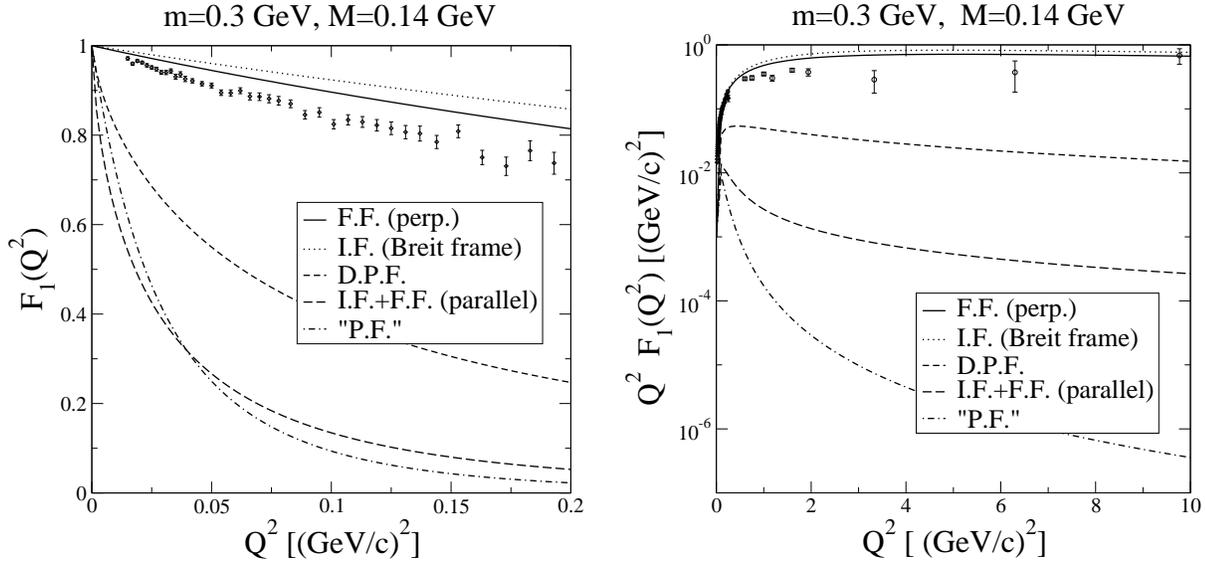

\includegraphics[width=0.48\textwidth]{figpis.eps}
 \hspace*{4mm}\includegraphics[width=0.48\textwidth]{figpiS.eps}
 \vspace*{-8mm}
\caption{Pion charge form factor in different forms 
of relativistic quantum mechanics.\label{fig1}}
\end{figure}

For a part, the present work was motivated by the drawbacks that 
an earlier point form  evidences for the form factors of strongly-bound 
systems calculated in the single-particle current approximation 
\cite{Desplanques:2001zw,Amghar:2002jx,Desplanques:2004sp}. Some results 
are presented in Fig. \ref{fig1} for the pion charge form factor. At high 
$Q^2$, the new point form (D.P.F.) shows a $Q^{-4}$ behavior, like the 
instant- and front-form results, while the earlier point form (``P.F.'') 
is providing a $Q^{-8}$ one. The change in the power law is largely due 
to the form of the velocity operator, Eq. (\ref{velocity}), which, 
containing some dependence on $\hat{u}$, makes less difficult to match the 
initial- and final-state momenta with those of the struck constituent. 
At low  $Q^2$, the new point form does better than the earlier one but the
improvement is not impressive. More important however, the bad behavior is
shared by results obtained in the instant and front forms with parallel
kinematics (I.F.+F.F.(parallel)) \cite{Desplanques:2004sp}. 
All of them evidence a charge squared radius scaling like the inverse 
of the  squared mass of the system. In comparison, the standard instant and 
front forms (I.F. (Breit frame) and F.F. (perp.) in Fig. \ref{fig1}) 
do well.

Lorentz invariance of form factors is often considered as an important criterion
for validating an approach. With this respect, the point form is to be favored as
it fulfills this property. It however recently appeared that the approaches that
give bad results above are strongly violating another important symmetry: 
Poincar\'e space-time  translation invariance \cite{Desplanques:2004sp}. 
Contrary to Lorentz invariance, this symmetry cannot be checked by looking at
form factors in a different frame. Instead, one could check relations such as:
\begin{equation}
\langle\Big[ P^{\mu}\;,\; J^{\nu}(x)\Big] \rangle = 
-i\langle \partial^{\mu}\,J^{\nu}(x) \rangle.
\end{equation}
\clearpage
This relation \cite{Lev:1993} cannot be verified exactly at the operator 
level in RQM approaches with a single-particle current but one can require 
it is verified, at least, at the matrix-element level. With this respect, 
what is an advantage for the point-form approach becomes a disadvantage 
as there is no frame where one can minimize the effect of a violation 
of the above relation (a factor 2-3 for the nucleon and 
roughly 6 for the pion). On the contrary, in the instant and front forms, 
one can consider different frames. It turns out that the instant- 
and front-form results for a perpendicular kinematics (standard  ones) 
verify the above equality while those for 
a parallel kinematics do badly, similarly to the point-form case. 
It thus appears that Poincar\'e space-time  translation invariance 
could be more important than the Lorentz one and that the intrinsic 
Lorentz covariance of the point-form approach is not so much  an advantage 
than what could be {\it a priori} expected.

\end{document}